# From annual to quarterly data: challenges and strategies in the estimation of Italian General Government Compensation of employees

Sara Cannavacciuolo, Maria Saiz, Maria Liviana Mattonetti [a]

[a] Istat, Italian National Institute of Statistics

**Abstract**

This paper addresses the methodology for the quarterly estimation of Compensation of Employees paid by the General Government (GG) sector, in accordance with the European System of Accounts (ESA 2010). Due to the limited high-frequency data availability and the need to guarantee the consistency with annual constraints, quarterly estimation relies on indirect temporal disaggregation techniques. These methods use specific infra-annual indicators as proxies for the variables being estimated. The specific case of the quarterly estimation of Compensation of employees presents several additional challenges. Firstly, the information provided by the sources, based on cash or legal-accrual data, is elaborated to define indicators which respect the accrual ESA 2010 principle as the annual estimates, based on more compliant data sources such as final budgets of public entities. Secondly, at a quarterly level the extraordinary events - such as the recording of delayed collective bargaining agreements which result in arrears - have a strong impact on quarterly indicators, whereas their effect is mitigated at annual level. To attribute these flows to the period when the work is performed, multi-source data harmonization techniques are employed. Thirdly, to accurately reflect intra-annual dynamics, information is collected for specific groups of GG entities (e.g., regions and provinces) and aggregated into ESA 2010 GG sub-sectors (Central Government, Local Government, Social Security Funds) leading to three specific estimates. To validate temporal disaggregation models and ensure methodological rigor and data quality, statistical tests are applied throughout the process. The results confirm the effectiveness of this methodology in providing accurate and timely quarterly estimates of Compensation of employees for the GG sector, thereby supporting reliable short-term economic analysis and policy making.

## 1. Introduction

The present paper addresses the methodology for the quarterly estimation of Compensation of employees paid by the General Government (GG) sector, in accordance with the European System of Accounts (ESA 2010) (Eurostat, 2013).[1]

Given the need of producing short-term estimates that are temporally consistent and aligned with official annual statistics, in accordance with Eurostat (2018), and considering the limited availability of sufficiently comprehensive high-frequency sources, indirect temporal disaggregation techniques are adopted. These methods allow for the estimation of quarterly series from lower-frequency data, using specific infra-annual indicators as *proxies* of the variables to be estimated.

---
[1] Opinion expressed in this paper are those of the authors and do not constitute policy of Istat.

In the challenging selection of high-quality indicators, a multi-sources approach is used, in order to account the various patterns of units included in GG sector. The information provided by Ministry of Economy and Finance (MEF) and Ministry of Health is adjusted to comply with the ESA 2010 accrual principle, with a specific treatment for extraordinary events as the systematic delayed implementation of collective wage agreements for civil servants. The resulting indicators are used to obtain quarterly estimates, through static regression models (Chow & Lin, 1971, Fernández, 1981, Litterman, 1983). Final estimates are then validated revealing their efficiency and ability in reflecting the short-term fluctuation of Compensation of employees.

The rest of the paper is organized as follows: Section 2 presents the methodological framework and Eurostat requirements, followed by the description of the procedure for deriving the indicators for quarterly estimates of Compensation of employees. Section 3 reports the results of the temporal disaggregation methods, validated through a large set of diagnostics.

## 2. Sources and indicators for temporal disaggregation methods

Among the literature on temporal disaggregation methods and benchmarking, procedures based on *proxy* variables have been the most popular and appropriately used, since they make use of relevant economic and statistical information. However, the resulting estimates depend critically on the choice of high-frequency indicators. As Nasse (1973) noted, these methods rest on an implicit assumption: the statistical relationship observed at a low-frequency between the *proxy* and the annual constraint also holds at higher frequencies (see Pavía-Miralles, 2010). According to Eurostat (2018), several key characteristics should be considered when selecting quarterly indicators. They should be timely and aligned with the reference period of the low-frequency target variable (i.e., annual constraint). Indicators should reflect the short-term fluctuation of the target variable while preserving a high correlation when aggregated to the lower frequency. In addition, indicators must be regular, exhibit a limited volatility and be conceptually consistent with the target variable.

With the aim to better reproduce the specific quarterly pattern of each GG sub-sector, the assessment of information available by GG sub-sector becomes crucial in the selection of suitable high-frequency indicators. ESA 2010 defines four GG sub-sectors: Central Government (CG), State Government - not applicable in countries without a federal structure, such as Italy, Local Government (LG), and Social Security Funds (SSF). The Italian framework of Quarterly non-financial accounts for GG, the estimation of Compensation of employees is based on a multi-source approach. More precisely, MEF provides quarterly data for the main central budgetary unit, the State, as well as for most of LG sector units, including Regions, Provinces, Municipalities and Universities, and for units within the SSF sub-sector. For the State, the information is on a legal-accrual basis (commitments)[2], whereas for the other units, information is available on a cash basis. In addition, the Ministry of Health supplies profit and loss account data for Local Health Units (LHU), included in LG sub-sector. All the sources cover the needed reference period and can be considered timely[3].

---

[2] A legal-accrual (commitment) refers to the act by which a public administration allocates part of a budget appropriation to cover a specific expenditure. It arises when a legal obligation is undertaken with a third party (e.g., when a contract is signed), regardless of when the actual payment will occur. Its purpose is to ensure that sufficient resources are available and to prevent expenditures from exceeding the amounts authorized by the budget. By contrast, under the accrual principle, costs and revenues are recorded when they are incurred, regardless of the timing of the payment or the budget commitment.

[3] The availability of sources differs among units: indicating with t the reference period, State data are supplied at t+20 days. MEF data on a cash basis are provided twice (i.e., t+20 and t+50), and finally, LHU data are available at t+50.

Since ESA 2010 requires that Compensation of employees must be recorded when the related work is performed (accrual principle), quarterly data are adjusted accordingly. In case of extraordinary events, additional information might be needed to refine the quarterly indicators series. A notable example concerns the delayed implementation of collective wage agreements for civil servants, which generates arrears. These arrears appear as irregular peaks in the indicator time series and are typically reflected in the quarterly data of Compensation of employees at the time of the actual disbursement (for cash data), or when the contract is signed (for legal-accrual). This phenomenon is documented by Agency for the Negotiating Representation of Public Administrations (ARAN), which publishes information on timing and type of contract signed, and is quantified by MEF, which provides annual cash and budgetary allocations for contract renewals. To comply with ESA 2010, quarterly sources are adjusted[4] to reallocate the wage increases to the appropriate reference period by: (i) netting arrears; (ii) imputing the accrued amounts in the quarters in which the agreements were delayed. The adjustment method depends on the recording principle underlying each source. For cash and legal-accrual data, the MEF's annual cash and budgetary allocations are distributed by using appropriate quarterly weights, and then incorporated into the quarterly sources. For LHU data, based on profit and loss accounts, the amounts of arrears are available separately as provisions, that can be reallocated to the relevant quarter. We applied this approach to develop indicators for GG sub-sector by considering contractual categories.

In this paper, we present the results and analysis by GG sub-sectors, beginning with the consistency between indicators and the low-frequency target variable, and then turning to the quarterly outcomes. Figure 4 shows the CG sub-sector quarterly indicator based on State data, which represents more than 92% of the Compensations of employees of the sub-sector. Over the period 1999-2024, the correlation between the annualized indicator and the annual constraint is 0.996 and the average coverage rate is 89.6%. Figure 5 presents the LG indicator constructed as the aggregation of the specific quarterly information corresponding to Regions, Provinces, Municipalities, Universities and LHU. This group of units represents more than 94% of LG sub-sector. Over years 1999-2024, the correlation between the annualized indicator and the annual constraint is 0.998 and the average coverage rate is 95.8%. Figure 6 shows the SSF quarterly indicator series. The correlation between the annualized indicator and the annual constraint is 0.953 and the average coverage rate is around 98.3%.

Figure 1-3) show that the indicators for each GG sub-sector are effective and precise in matching the annual National accounts constraints.

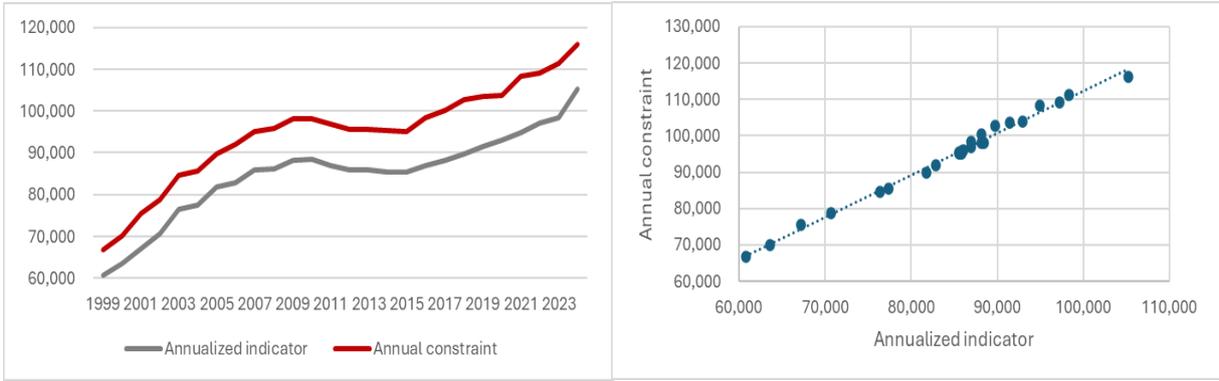

*Figure 1. Compensations of employees: annualized indicator and annual constraint series – CG sub-sector. 1999-2024. Millions of euros.*

---

[4] This recording method, agreed with Eurostat, has been implemented from 2016 onwards, in both the annual and quarterly National accounts.

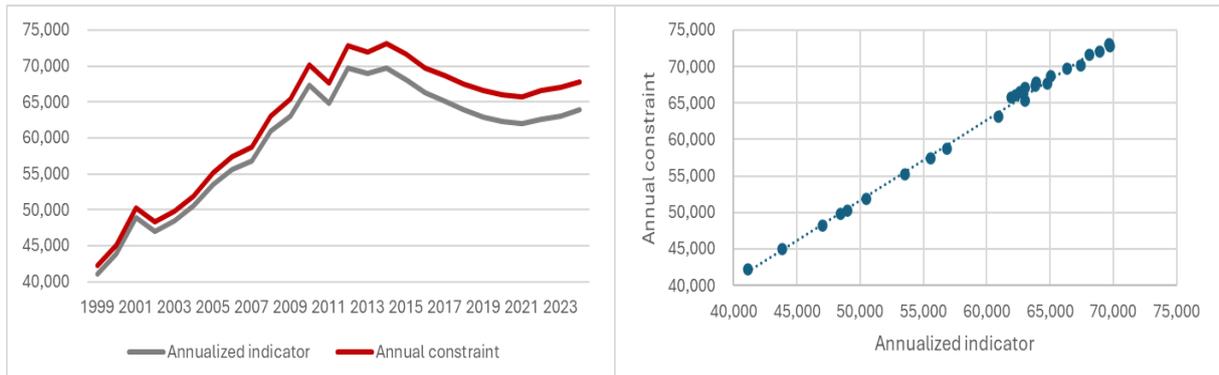

*Figure 2. Compensations of employees: annualized indicator and annual constraint series – LG sub-sector. 1999-2024. Millions of euros.*

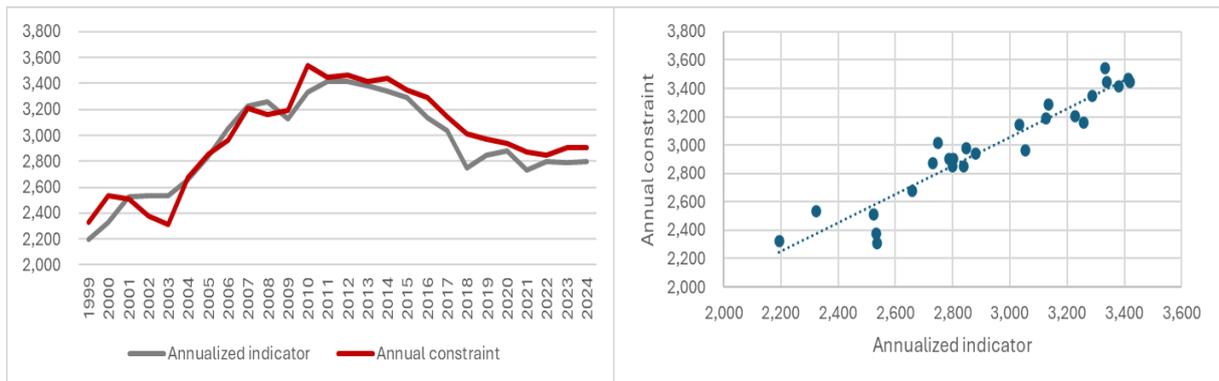

*Figure 3. Compensations of employees: annualized indicator and annual constraint series – SSF sub-sector. 1999-2024. Millions of euros.*

## 3. Results

To ensure temporal consistency in the compilation of GG time series, temporal disaggregation methods are applied according to guidelines provided by Eurostat (2018). A statistical approach is adopted, considering static regression methods notably Chow & Lin (1971), Fernández (1981) and Litterman (1983). These methods fit models with high frequency indicator series regressed on the low-frequency series (annual constraint) to be disaggregated. The methods differ in the proposed models for the structure of residuals and are selected using a large set of descriptive and parametric criteria including graphical analysis, also account for computational complexity. For a detailed description of implemented models and diagnostics see Moauro & Bisio (2016). The procedure is implemented via SAS/IML packages that enables the validation of results through statistical tests. As presented in Table 1, the quarterly estimations of Compensation of employees for CG and LG sub-sectors are carried out using the Fernández model, due to the strong reliability of the selected indicators in capturing short-term fluctuations of the target variable. On the other hand, the Chow-Lin model is selected for the estimation of SSF sub-sector, owing to its suitability for the characteristics of the series. In addition, external variables are incorporated into the models to account for relevant economic events and the employment pattern.

Temporal disaggregation results are validated through an extensive set of diagnostics, including those reported in Table 2 and Table 3. The selected models demonstrate goodness-of-fit, with Gaussian, homoscedastic and non-autocorrelated residuals. Figure 4-6) compare disaggregated results to their respective sub-sector indicators, confirming the indicators effectiveness in reproducing short-term fluctuations of the target variable. Given the presence of the seasonal component, as observable in the pattern of the time series presented in Figures (4-6), quarterly indicators are also seasonally adjusted (SA). The strategy implemented is that described in Mattonetti & Saiz (2025), enabling the application of temporal disaggregation tecniques on the SA indicator and producing quarterly estimates free from seasonal component.

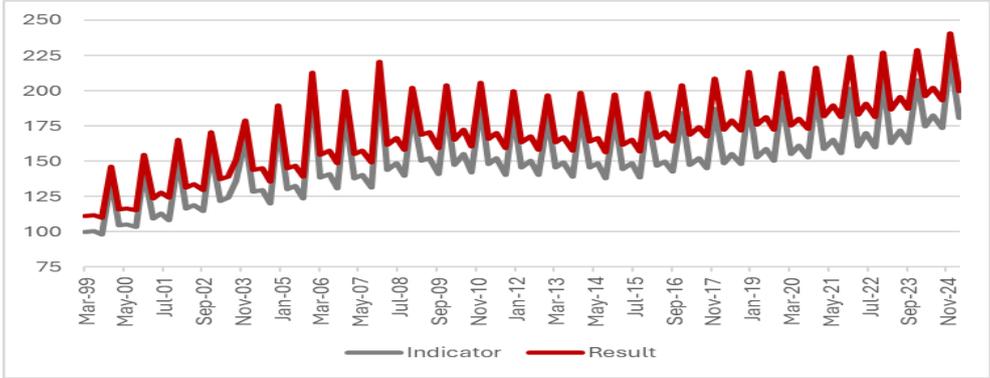

*Figure 4. Compensations of employees: quarterly indicator and disaggregated series – CG sub-sector. 1999Q1-2025Q1. Not seasonally adjusted data. Transformed data: 1999Q1 indicator series=100.*

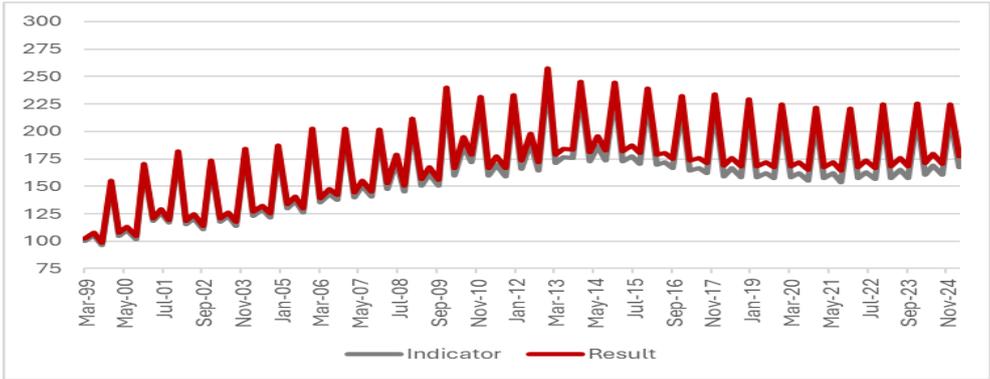

*Figure 5. Compensations of employees: quarterly indicator and disaggregated series – LG sub-sector. 1999Q1-2025Q1. Not seasonally adjusted data. Transformed data: 1999Q1 indicator series=100.*

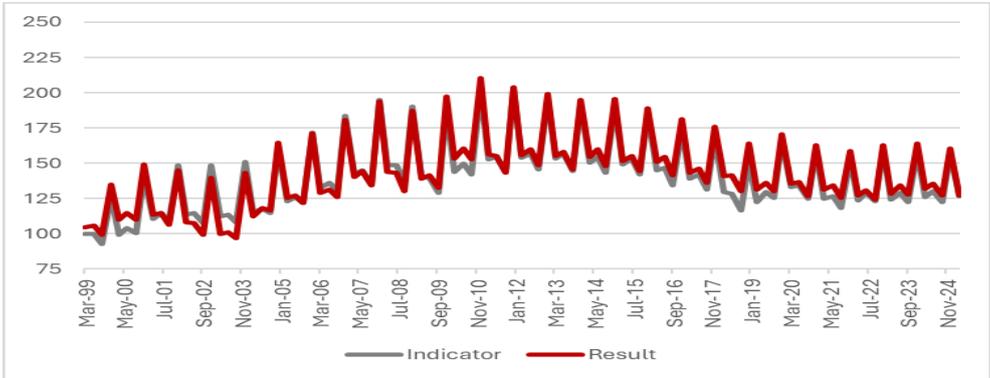

*Figure 6. Compensations of employees: quarterly indicator and disaggregated series – SSF sub-sector. 1999Q1-2025Q1. Not seasonally adjusted data. Transformed data: 1999Q1 indicator series=100.*

|  | Coefficient estimate | Standard error | t-statistic | p-value |
|---|---|---|---|---|
| CG sub-sector – Fernández model | | | | |
| $\widehat{y_{it}} = \hat{c} + \widehat{\beta_1} i_{it} + \widehat{\beta_2} SD(2012,2015) + \widehat{\beta_3} SD(2021,2024)$ | | | | |
| $\hat{c}$ | 2478.600 | 760.370 | 3.259 | 0.003 |
| $\widehat{\beta_1}$ | 0.917 | 0.057 | 16.110 | 1E-14 |
| $\widehat{\beta_2}$ | -300.800 | 135.250 | -2.224 | 0.035 |
| $\widehat{\beta_3}$ | 830.480 | 190.860 | 4.351 | 2E-4 |
| LG sub-sector – Fernández model | | | | |
| $\widehat{y_{it}} = \hat{c} + \widehat{\beta_1} i_{it} + \widehat{\beta_2} SD(2012,2020)$ | | | | |
| $\hat{c}$ | 70.026 | 104.940 | 0.667 | 0.510 |
| $\widehat{\beta_1}$ | 1.021 | 0.010 | 102.360 | 0.000 |
| $\widehat{\beta_2}$ | -48.184 | 22.398 | -2.151 | 0.041 |
| SSF sub-sector – Chow-Lin model | | | | |
| $\widehat{y_{it}} = \widehat{\beta_1} i_{it}$ | | | | |
| $\widehat{\beta_1}$ | 1.022 | 0.010 | 103.080 | 0.000 |

*Table 1. Compensation of employees: Temporal disaggregation model summary by GG sub-sector.*

| Sub-sector | ρ | $R^2$ | Standard error | F-test | Durbin-Watson test | Jarque-Bera test | H test | Ljung-Box Q-test |
|---|---|---|---|---|---|---|---|---|
| CG | 1 | 0.997 | 135.844 | 1478.675 (0.000) | 2.760 | 0.927 (0.629) | 0.064 (1) | 13.424 (0.062) |
| LG | 1 | 1 | 23.734 | 30769.556 (0.000) | 2.005 | 1.456 (0.483) | 0.011 (1) | 6.830 (0.447) |
| SSF | 0.731 | 0.972 | 26.048 | 90.670 (0.000) | 2.067 | 0.272 (0.873) | 1.459 (0.148) | 4.060 (0.773) |

*Table 2. Compensation of employees: Temporal disaggregation models standard diagnostics by GG sub-sector. P-values in parentheses.*

| Sub-sector | Correlation among QoPQ(%)* | Correlation among QoSQ(%)** | RMSE among QoPQ(%)* | RMSE among QoSQ(%)** | Max discrepancy among QoPQ(%)* | Max discrepancy among QoSQ(%)** |
|---|---|---|---|---|---|---|
| CG | 1 | 0.965 | 3.962 | 1.179 | -11.377 | -3.303 |
| LG | 1 | 0.999 | 0.600 | 0.240 | -1.607 | 0.586 |
| SSF | 0.997 | 0.603 | 1.671 | 4.563 | 8.708 | 14.620 |

*QoPQ(%) stays for Quarter On Previous Quarter Percent Changes.

** QoSQ(%) stays for Quarter On Same Quarter A Year Ago Percent Changes.

*Table 3. Compensation of employees: Temporal disaggregation statistics calculated over the indicator series and the disaggregated series, presented by GG sub-sector.*